\documentclass[14pt, twocolumn]{article}

\usepackage{blindtext}
\usepackage{multicol}
\usepackage{amsmath}
\usepackage{graphicx}

\usepackage{graphics}

\usepackage{xcolor}

\usepackage[T2A]{fontenc}
\usepackage[utf8]{inputenc}

\usepackage{hyperref}
\usepackage{indentfirst}
\usepackage[labelfont=bf]{caption}
\usepackage{authblk}
\usepackage{graphicx}

\usepackage{cite}

\usepackage{geometry}
\begin{document}

\title{Numerical study of GCR proton transport}
\author[1]{Yurovsky V.O.}
\author[1]{Peryatinskaya A. I.}
\author[2]{Kudryashov I.A.}
\affil[1]{Faculty of Physics, MSU, Moscow, Russia}
\affil[2]{SINP MSU, Moscow, Russia}
\affil[ ] {e-mails: \href{mailto:ilya.kudryashov.85@gmail.com}{ilya.kudryashov.85@gmail.com}, \href{mailto:yrovskyvladimir@gmail.com}{yrovskyvladimir@gmail.com} }

\date{}

\maketitle

\begin{abstract}
In the paper numerical investigation of the propagation of galactic cosmic rays in a model magnetic field is discussed. The magnetic field is modeled as a composition of an isotropic turbulent field with the Kolmogorov turbulence spectrum and a regular constant field. The dependence of the diffusion tensor components on the particle energy is studied.
\end{abstract}

\section{Introduction}

The recent development of experimental techniques in the field of cosmic ray physics has made it possible not only to refine the structure of the well-known knee in the rigidity spectrum of cosmic rays range of 1-10 PV, but also to discover several new features. Firstly, it is a universal inhomogeneity in the energy spectra of various hadron components of cosmic rays in the range of rigidities 1 TV - 100 TV, discovered in the NUCLEON experiment\cite{nucleon}, and confirmed by other experiments, such as CALET\cite{calet} and DAMPE\cite{dampe_exp}. Secondly, it is a sharp change in the amplitude and phase of the dipole anisotropy of galactic cosmic rays in the energy range 100 TeV - 1 PeV\cite{dipole_phase}. To explain these phenomena one needs a detailed understanding of the mechanisms of CR transport in a wide energy range from hundreds of GeV to hundreds of PeV in a realistic magnetic field.

Existing models of the magnetic field predict a large range of turbulence - from 100 astronomical units to 100 parsecs (for more details, see Chapter 2). To correctly solve the problem of propagation of galactic cosmic rays in the interstellar medium, it is necessary to take into account the entire spatial range of magnetic field inhomogeneities, which required writing our own code.

Many studies have been devoted to the problem of the transport of relativistic charged particles. Numerical approaches are most often used\cite{numerical1, numerical2, numerical3, numerical4, numerical5}, there are also attempts to theoretically describe the mechanisms of transport\cite{teor1, teor2, teor3}, but a theory that has predictive power for transport in isotropic turbulence has not yet been built. 

To calculate the transport of relativistic charged particles in a realistic turbulence range, we performed numerical simulation using a specially written software package.

The article is structured as follows. In Chapter 2, we describe our transport model: an algorithm for constructing a magnetic field and calculating the transport of particles in it. Chapter 3 describes the calculation of diffusion coefficients and introduces the concept of diffusion coefficient for magnetic field lines. Chapter 4 presents and discusses the results obtained. Chapter 5 contains the conclusion.




\section{Transport model}

\subsection{Magnetic field}

In this work the transport of particles is studied in two magnetic fields types.

The first type is a random isotropic turbulent field. This random field $\boldsymbol b$ is represented by the sum of a finite number of several randomly generated modes:

\begin{equation}
\boldsymbol b = \sum_n{A_n \boldsymbol P_n cos(\boldsymbol r \boldsymbol k_n + \phi_n) }
\end{equation}

In this work $N = 500$. The same magnetic field model was studied, for example, in \cite{magnetic_field}. The modes are generated with a logarithmic step in the spatial size from 100 a.u. up to 100 pc, the correlation length is 20 pc. The RMS value of the turbulent field is 6 \textmu G.

\begin{figure}[h]
    \includegraphics[width=\linewidth]{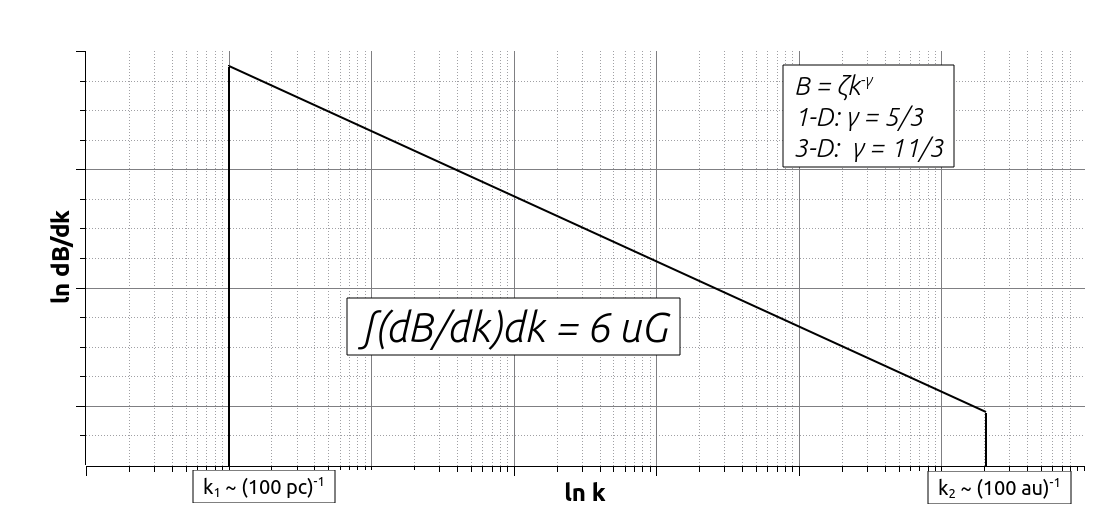}
    \caption{Spectrum of the turbulent component of the field}
    \label{fig:spectr}
\end{figure}

The spectrum of the field $b$ is shown in Fig.\ref{fig:spectr}

The second type of field is a sum of the first type turbulent field $\boldsymbol b$ with a constant field $\boldsymbol B_0$. The value of this constant field is proportional to the RMS value of the turbulent field. In this work we study cases with $\xi = \frac{B_0}{RMS(b)} = 0; 0.25; 0.5; 1 $. Also, in out work we assume, that $\boldsymbol{B_0}$ is directed along $z$ axis.

\begin{equation}
\boldsymbol B = \boldsymbol B_0 + \boldsymbol b,
\end{equation} 

In contrast to CRPROPA\cite{crpropa}, the field is not calculated in advance at set grid nodes, and then interpolated. Instead, it is dynamically calculated at each point of the trajectory, which makes it possible to take into account small-scale turbulence.

\subsection{Magnetic field generation check}

\begin{figure*}[t]
    \includegraphics[width=\textwidth]{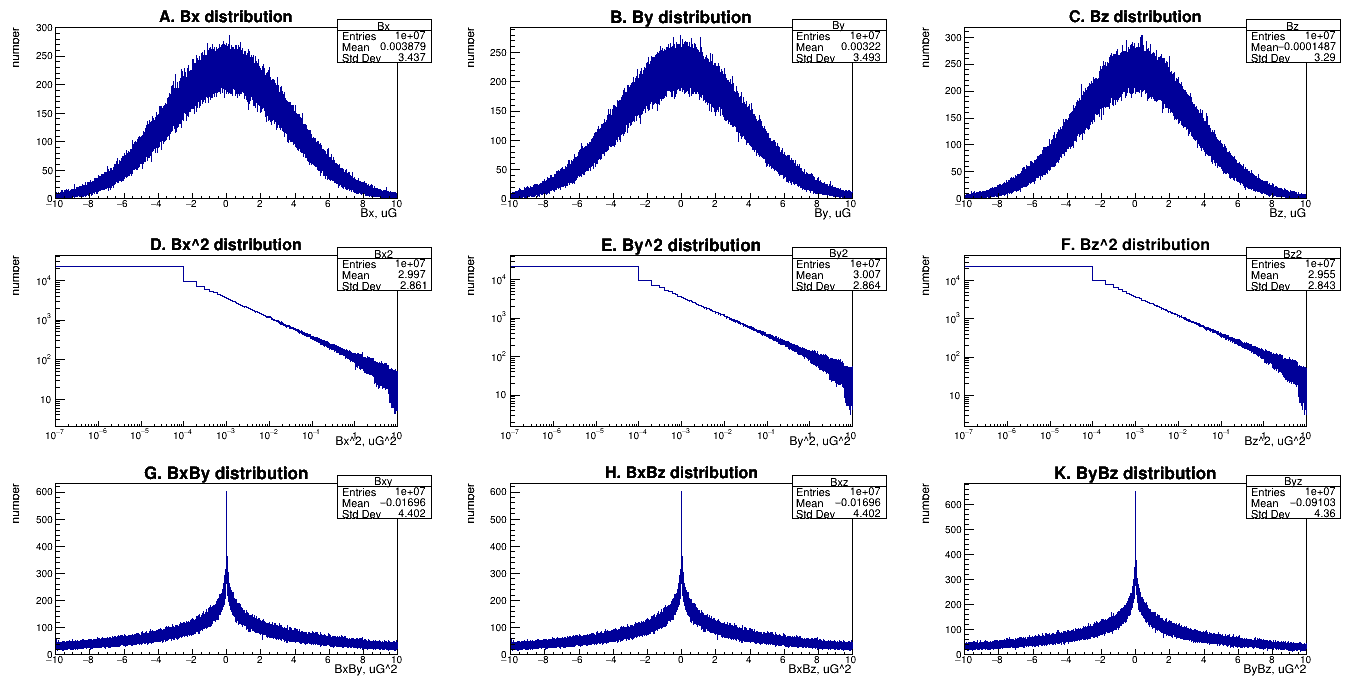}
    \caption{Histograms of magnetic field values. (A) $b_x$ (B) $b_y$ (C) $b_z$ (D) $b_x^2$ (E) $b_y^2$ (F) $b_z^2$ (G) $b_xb_y$ (H) $b_xb_z$ (K) $b_yb_z$}
    \label{fig:res1}
\end{figure*}

The method of turbulent field generation described above was tested as follows: histograms of $b_i$, $b_ib_j$ values were constructed for measured $10^7$ values at random points in the realisation of turbulent magnetic field. One of the results is shown on the figure \ref{fig:res1}.

This figure shows that the distribution of the components is normal. From the distributions of $b_ib_j$ on the same figure it can also be seen that individual axial components do not correlate.

\subsection{Propagation}
In this work the propagation of a particle in a magnetic field was described with the following equations:

\begin{equation}
\begin{cases}
\frac {d \boldsymbol r}{dt} = \boldsymbol v \\
\gamma m \frac{d\boldsymbol v}{dt} = q[\boldsymbol v \times \boldsymbol B],
\end{cases}
\label{eq1}
\end{equation}

which can be rewritten as:

\begin{equation}
\begin{cases}
\frac {d \boldsymbol r}{dt} = c \tilde{\boldsymbol v} \\
\frac{ d \tilde{\boldsymbol v}}{dt} = \frac{qc^2}{E} [\tilde{\boldsymbol v} \times \boldsymbol B],
\end{cases}
\end{equation}

where E is the particle energy, and $\tilde{\boldsymbol v}$ is a unit length vector along the particle's velocity.

These equations are solved using the Cash-Karp method, a membed of the Runge-Kutta family of the 4th order of accuracy.

Methods from the Runge-Kutta family are non-conservative. The modification of Newton's equations shown above is needed in order for the energy conservation law to be fulfilled in an explicit form. The time step was chosen empirically to ensure correct modeling of particle motion in a constant magnetic field. To check whether the step is small enough in the case of the model field, we ran the simulation with a step 10 times larger than the chosen one and the result was statistically indistinguishable from the initially chosen one. For example, for a particle energy of 100 TeV, steps of 100 AU were chosen ($\sim$ 1/30 gyroradius in constant magnetic field with magnitude $RMS(b)$).

\begin{equation}
dt = \frac{E \frac{au}{TeV}}{c}
\end{equation}

\section{Calculation of diffusion coefficients}
To calculate the diffusion coefficients we averaged trajectories of particles in three random realizations of the magnetic field with same spectrum for each energy value. Each run was a simulation of the propagation of 32 particles for a particle run of 800 kpc. Each attempt used a new random implementation of the turbulent field. The running diffusion coefficient is defined as

\begin{equation}
D_{ij} (l) = \frac{x_i(l)x_j(l)}{l/c}.
\end{equation}

\begin{figure}
    \includegraphics[trim={0 190px 0 150px},clip,width=\linewidth]{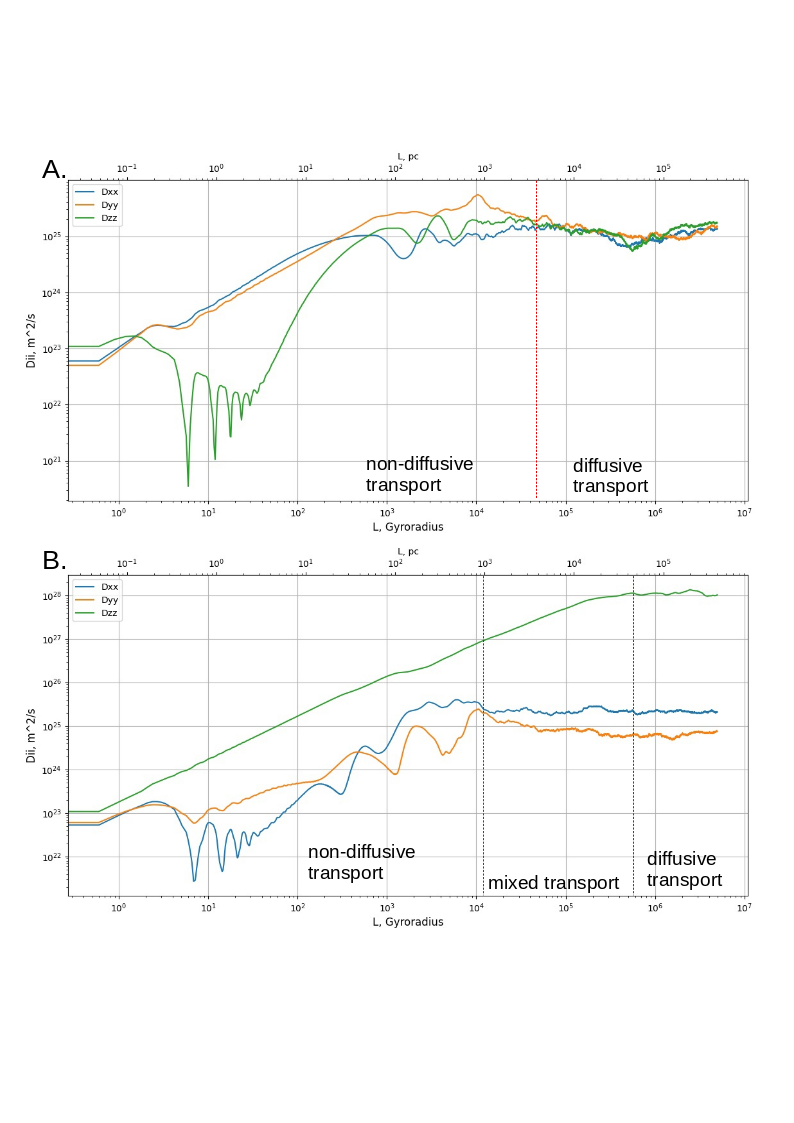}
    \caption{Transition to diffusive transport from ballistic. (A) Isotropic case with $\xi = 0$ (B) Anisotropic case with $\xi = 1$}
    \label{fig:drrL}
\end{figure}

From the results of the experiment, it can be seen(Fig. \ref{fig:drrL}) that for sufficiently large l it becomes constant, so that in the end we obtain simply the diffusion coefficient as 

\begin{equation}
D_{ij} = \lim_{l\to\infty} D_{ij}(l).
\end{equation}

\subsection{Diffusion coefficient for magnetic field lines}

In turbulent field magnetic field lines are are very mixed up, so we can say about diffusion of field lines.
For the case of isotropic turbulence (without a regular field), the field lines diffusion coefficient was calculated as

\begin{equation}
D^f_{ij} = \lim_{l\to\infty} \frac{x_i(l)x_j(l)}{l}.
\end{equation}

Magnetic field lines diffusion coefficient is important because, in the low-energy limit, the particles are completely captured in the field and the leading center approximation is applicable. In this case,  particles are mainly propagate along the field lines and this transport is faster, than diffusion. But macroscopic transport is still diffusive due to the diffusion of magnetic lines. So diffusion of magnetic filed lines should be in a sense a limit case for the particle diffusion.

\section{Results}

\subsection{Magnetic field line diffusion}
\begin{figure}[h]
    \centering
    \includegraphics[width=\linewidth]{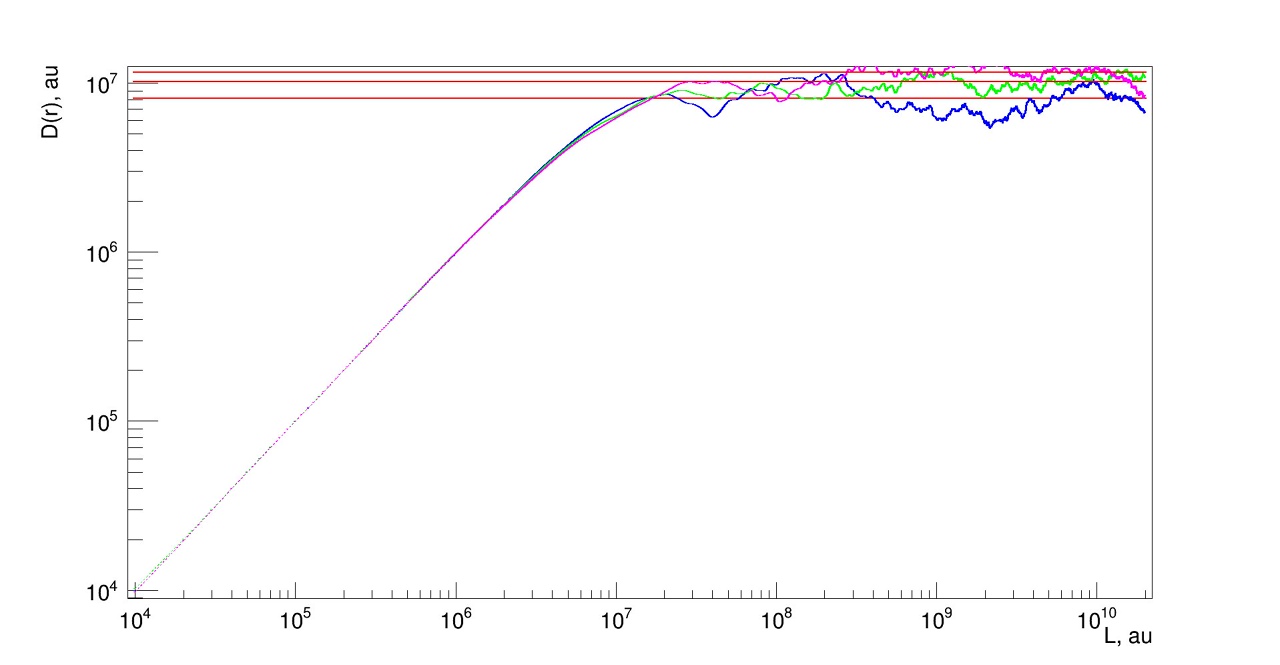}
    \caption{Values of diffusion coefficients of field lines for three realizations of the field}
    \label{fig:lines}
\end{figure}

The simulation results for three isotropic magnetic field realisations are shown in fig. \ref{fig:lines}.

The dimension of this coefficient is equal to the dimension of length. If one wants to compare this coefficient with the diffusion coefficient of particles, then coefficient for lines should be multiplied by $c/2$, where $c$ is the speed of light. This corresponds to the fact that the particles fly at the speed of light, and the angle between the speed and the direction of the field line is uniformly distributed. For our case of turbulent magnetic field we get that diffusion coefficient for magnetic field lines $\approx 10^{26} m^2/s$. 

\subsection{Isotropic turbulence}

\begin{figure}[h]
    \centering
    \includegraphics[width=\linewidth]{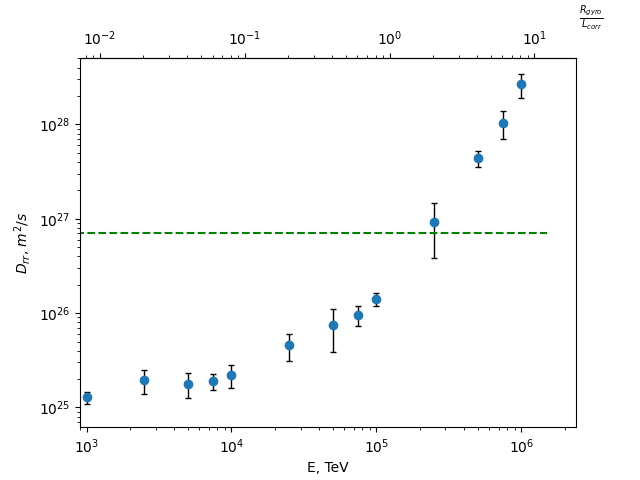}
    \caption{Averaged values of $Drr$ in the isotropic case. Green line shows diffusion coefficient of field lines}
    \label{fig:isoDrr}
\end{figure}

The energy dependence of the diffusion coefficient for the case of isotropic turbulence is shown on Fig. \ref{fig:isoDrr}.

\begin{figure}[h]
    \centering
    \includegraphics[trim={0 670px 0 0},clip,width=\linewidth]{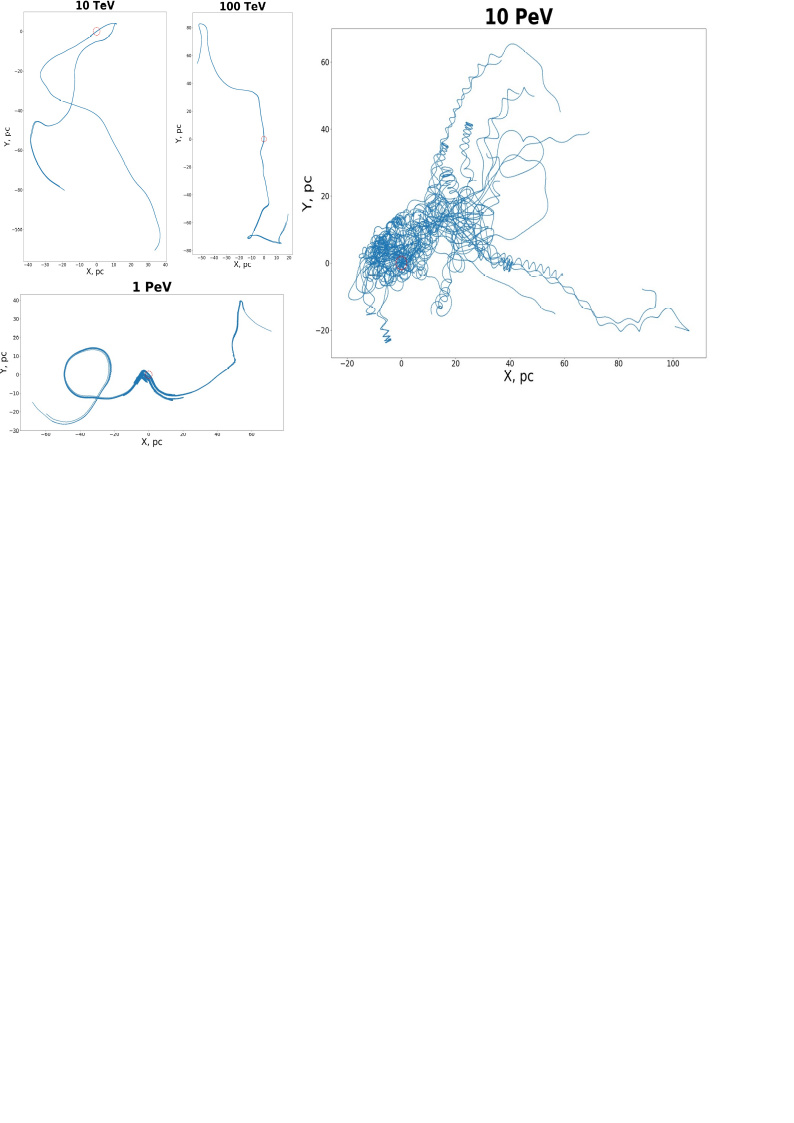}
    \caption{Trajectories of 32 particles with mileage 100 pc started from one point with velocities uniformly distributed over the sphere with different energies.}
    \label{fig:transport}
\end{figure}

It can be seen that $D_{rr}$ is constant for low energies, and then begins to increase as the energy increases. This can be explained as follows: when the gyroradius of a particle in the mean field is small compared to the characteristic size of inhomogenities in the magnetic field we can use the guiding center approximation. In this approximation the particles predominantly propagate along the field lines of force, and their spatial propagation is determined mainly by the diffusion of field lines, which will be separately studied in future papers. When the gyroradius of the particle in the mean field reaches the characteristic size of inhomogenities in the magnetic field, the particle diffusion regime begins. These considerations is illustrated at Fig. \ref{fig:transport}. It can be seen on this figure that transport pattern changes with increasing energy.

\begin{figure}[h]
    \centering
    \includegraphics[width=\linewidth]{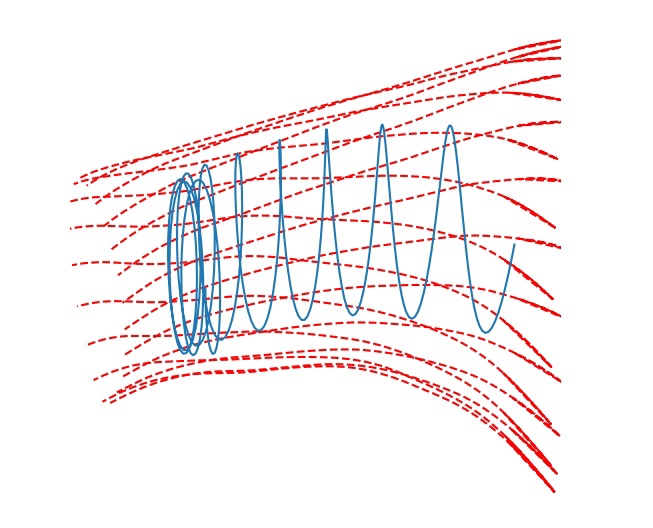}
    \caption{Example of magnetic trap. Red dashed lines are magnetic field lines, blue solid line is particle trajectory.}
    \label{fig:trap}
\end{figure}

If we compare the value of the coefficient of the field line with the value obtained from the dependence of the diffusion coefficient for particles in the case of an isotropic turbulent field, we will see that the coefficient for particles is an order of magnitude lower than for the field. This may be due to the influence of reflecttion on magnetic field inhomogenities. Example of such reflection is shown on the figure \ref{fig:trap}. It can be seen, that particle is reflected from the area of field line concentration. Such reflections can slow down transport of particles along field lines and decrease diffusion coefficient. This will also be studied in future works. 

\subsection{Anisotropic turbulence}
\begin{figure}[h]
    \centering
    \includegraphics[width=\linewidth]{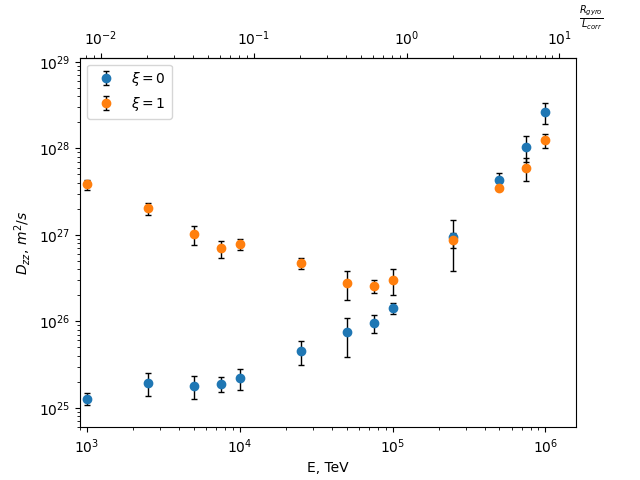}
    \caption{Average values of $Dzz$ for $\xi = 0, 1$}
    \label{fig:b01Dzz}
\end{figure}

Recall that we studied cases with $\xi = 0;0.25;0.5;1$. On fig. \ref{fig:b01Dzz} the energy dependencies of the diffusion coefficient for the cases $\xi = 0$ and $\xi = 1$ are shown. There is also a second horizontal axis on the chart. The values on it are $x = R(E) / L_{corr}$, where $R(E) = \frac{E}{qcb}$ is the gyroradius of the particle in the RMS(b), and $L_{corr}$ is the correlation length of the magnetic field.

\begin{figure}[h]
    \centering
    \includegraphics[width=\linewidth]{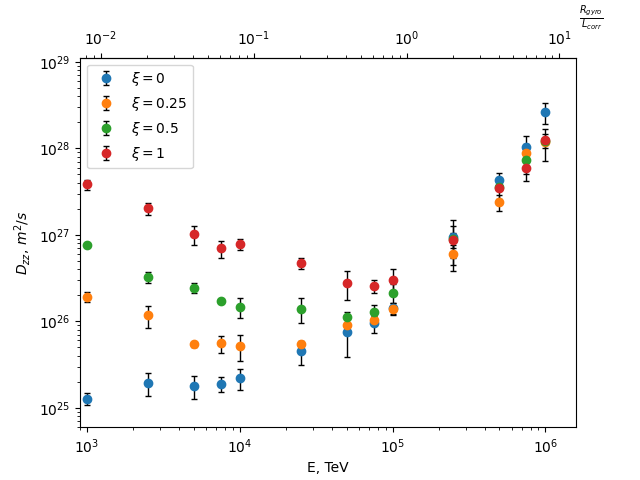}
    \caption{Average values of $Dzz$ for $\xi = 0, 0.25, 0.5, 1$}
    \label{fig:ballDzz}
\end{figure}

On fig. \ref{fig:ballDzz} the dependencies of $D_{zz}$ on energy are presented for all the studied cases of the $\xi$.

This figure shows a general trend: $D_{zz}$ first falls, and the larger the $\xi$, the steeper the slope. Then the graph $D_{zz}(E)$ merges the graph $D_{zz}(E)$ for the case $\xi = 0$ (at $\approx 100 PeV$), and this merging does not depend on the $\xi$ value. Similar behavior was also observed for other magnetic field models, for example, in \cite{same_res}.

\subsection{Perpendicular transport}
\begin{figure}[h]
    \centering
    \includegraphics[width=\linewidth]{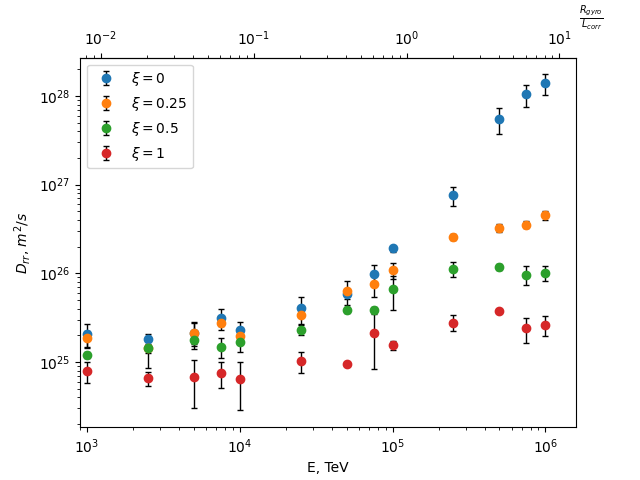}
    \caption{Average $Drr$}
    \label{fig:Drr}
\end{figure}

On fig. \ref{fig:Drr} the dependencies of the diffusion coefficient in the plane perpendicular to the direction of the constant magnetic field are shown, represented by $D_{rr} = \frac{Dxx+Dyy}{2}$. Division by 2 is needed for $D_{rr}$ to reflect the displacement projection onto some vector (in the XY plane), which, in turn, is necessary for the correct comparison of $Dzz$ and $Drr$ values.

It can be seen that the behavior of $D_{rr}$ is even more complicated than that of $Dzz$. For low and high energies it is constant (except for the $\xi = 0$ case), and only in a rather narrow energy range is there some increase. The smaller $\xi$, the larger value of $D_{rr}$ increase. The explanation of this phenomenon requires further research.

\subsection{Transport anisotropy}
\begin{figure}[h]
    \centering
    \includegraphics[width=\linewidth]{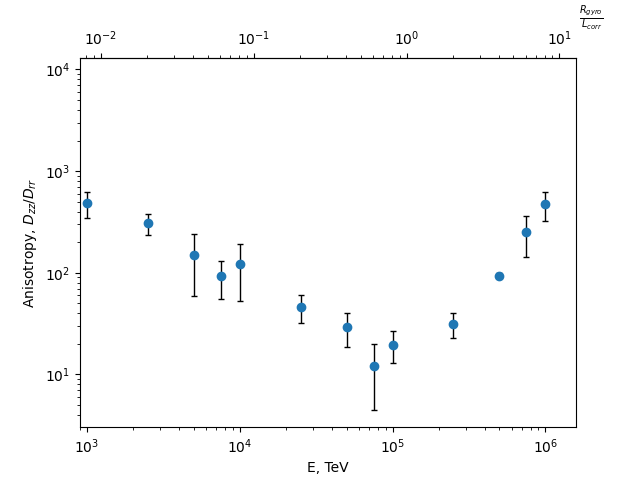}
    \caption{Energy dependence of diffusion tensor anisotropy at $\xi = 1$}
    \label{fig:aniso}
\end{figure}

On fig. \ref{fig:aniso} the energy dependence of the diffusion tensor anisotropy $a = \frac{D_{zz}}{D_{rr}}$ for the $\xi = 1$ case is shown.

This figure shows, that the anisotropy rises with the $\xi$. It can also be seen that the anisotropy $a$ is $> 10$ in presence of a constant field, which means that, diffusion is anisotropic. The diffusion tensor is anisotropic at any energy and for the studied cases of the $\xi$ ratios considered, but the observed anisotropy at a point is determined by the cosmic ray concentration gradient only in those cases where the particle does not move along the field line. If the particle is captured by the field, then the observed anisotropy at the point shows the direction of the field line passing through the point of observation. This issue will be considered in more details in the future works.

\section{Conclusion}

The authors have constructed a numerical model for the propagation of relativistic charged particles in a synthetic magnetic field, which makes it possible to realize a large range of magnetic field turbulence. Using this model, the components of the diffusion tensor of particles were calculated for various field configurations. The diffusion coefficients for the field lines were also calculated. 

The calculations carried out using this model showed a complex behavior of the transport coefficients with increasing mileage. At short mileages, transport has a superdiffusive character, and only at sufficiently large mileages transforms into a diffusive one. It should be noted that for the case of the presence of a regular field component, mileage sufficient to reach the diffusive transport regime for the transverse and longitudinal diffusion coefficients is different.

The dependencies of the diffusion coefficients on the particle energy are obtained. It is shown that the longitudinal and transverse diffusion coefficients are characterized by complex behavior. We assume that this behavior can be explained by changing in the propagation mechanism. Relatively low-energy particles propagate along the magnetic field lines, and the observed diffusive particle transport is a combination of one-dimensional particle transport along the field lines and diffusion of the field lines themselves. The transport of a high-energy particle has the character of three-dimensional anisotropic diffusion and is not correlated with the transport of field lines.

\bibliographystyle{unsrt}
\bibliography{refs.bib}


\end{document}